\documentclass[superscriptaddress,twocolumn,nofootinbib,prl,amsmath]{revtex4}
\usepackage{times,color}
\usepackage{graphicx}
\usepackage{color}
\usepackage{amsmath}
\usepackage{amssymb}

\newcommand{\beq}{\begin{equation}}
\newcommand{\eeq}{\end{equation}}
\newcommand{\ket} [1] {\vert#1\rangle}

\newcommand{\ra}{{\ket{r}}_A}
\newcommand{\rb}{{\ket{r}}_B}
\newcommand{\la}{{\ket{\ell}}_A}
\newcommand{\lb}{{\ket{\ell}}_B}

\begin{document}
\title{Phase control of a longitudinal momentum entangled photon state by \\a deformable membrane mirror}

\author{Cristian Bonato}
\address{CNR-INFM LUXOR, Department of Information Engineering, University of Padova, Padova, Italy}
\address{Huygens Laboratory, Leiden University, P.O. Box 9504, 2300 RA Leiden, the Netherlands}
\author{Stefano Bonora}
\homepage{http://www.padova.infm.it/luxor/}\address{CNR-INFM LUXOR, Department of Information Engineering, University of Padova, Padova, Italy}
\author{Andrea Chiuri}
\address{Dipartimento di Fisica, Universit\`{a} Sapienza di Roma,
Roma 00185, Italy}
\author{Paolo Mataloni}
\homepage{http://quantumoptics.phys.uniroma1.it/}
\address{Dipartimento di Fisica,
Universit\`{a} Sapienza di Roma,
Roma 00185, Italy}
\address{Istituto Nazionale di Ottica Applicata (INOA-CNR), L.go E. Fermi 6, 50125 Florence, Italy}
\author{Giorgio Milani}
\homepage{http://quantumoptics.phys.uniroma1.it/}
\address{Dipartimento di Fisica, Universit\`{a} Sapienza di Roma,
Roma 00185, Italy}
\author{Giuseppe Vallone}
\homepage{http://quantumoptics.phys.uniroma1.it/}
\address{Museo Storico della Fisica e Centro Studi e Ricerche ``Enrico Fermi'', Via Panisperna 89/A, Compendio del Viminale, Roma 00184, Italy}
\address{Dipartimento di Fisica, Universit\`{a} Sapienza di Roma,
Roma 00185, Italy}
\author{Paolo Villoresi}
\homepage{http://www.padova.infm.it/luxor/}\address{CNR-INFM LUXOR, Department of Information Engineering, University of Padova, Padova, Italy}

\date{\today}

\begin{abstract}
We propose a paradigmatic demonstration of the potentialities of a deformable mirror for closed-loop control of a two-photon momentum-entangled state, subject to phase fluctuations. A custom-made membrane mirror is used to set a relative phase shift between the arms of an interferometric apparatus. The control algorithm estimates the phase of the quantum state, by measurements of the coincidence events at the output ports of the interferometer, and uses the measurements results to provide a feedback signal to the deformable mirror. Stabilization of the coincidence rate to within $1.5$ standard deviation of the Poissonian noise is demonstrated over $2000$ seconds.
\end{abstract}

\maketitle

Adaptive optimization of experimental parameters is an extremely powerful tool for researchers. In optics, higher-order material dispersion, broadband phase matching conditions in inhomogeneous media as well as the nonlinear and thermal deformation to the pulse wavefront pose solid difficulties to the achievement of the optimal conditions for the processes under study.  The adaptive approach, i.e. the use of suitable devices that may circumvent the limits of conventional components by adapting their shape, has paved the way to several breakthroughs in the generation of quantum states via nonlinear optical phenomena as well as in the transformation in time and frequency of laser pulses up to the single optical cycle regime. 

Adaptive optics was developed with the idea to act on portions of an optical beam to correct aberrations. The action is driven by the direct measure of the alterations, as explored over a century ago in the case of astronomical instrumentation by {J. A. Hartmann} \cite{hart00zin}. Indeed, the initial applications of adaptive optical devices were in astronomy, due to the possibility of correcting the wavefront of a beam gathered by a telescope, thus compensating the degradation due to atmospheric turbulence \cite{tysoSPIE}. From these initial applications, adaptive optics has spread into different fields, like laser physics \cite{zeek00opl, brid08opl},
biomedical imaging and vision \cite{knut03ope}. Deformable mirrors and, in particular, membrane mirrors appear to be particularly interesting due to their low losses, insensitivity to chromatism and large dynamics. Furthermore, these mirrors are cheap and are characterized by a low power consumption.

More recently, deformable mirror have been used in a few seminal experiments in quantum optics.
In a first experiment \cite{abou02prl}, a segmented michroelectromechanical micromirror was used to demonstrate
that a coherent image of a pure phase object can be obtained using the interbeam coherence of a
pair of spatially incoherent entangled photon beams. In a second experiment \cite{bona08prl},
a membrane deformable mirror was used to demonstrate the even-order aberration cancelation effect
in quantum interferometry. The adaptive mirror allowed a precise and clean implementation of selected
optical aberrations, so that it was possible to show experimentally that the second-order correlation
function for a pair of entangled photons is sensitive only to odd-order aberrations. In both experiments,
however, the deformable mirror was used as a static device, where a specific shape was dialed and then kept
fixed for the duration of the experiment.

In the present work we give a paradigmatic example of the potentialities of a membrane mirror for classical
closed-loop control of a two-qubits entangled optical state. In particular, we have used it to stabilize the
phase of a two photon state entangled in the degree of freedom of longitudinal momentum {\cite{rari90prl}},
subject to random fluctuations. The experiment has been realized by adopting a simplified version of the
apparatus recently introduced to demonstrate the entanglement of two photons in many spatial optical
modes(multipath entanglement \cite{ross09prl}). A stream of momentum-entangled photon pairs propagates through
an interferometric optical system in which random optical path length instabilities result in fluctuations
of the relative phase of the quantum superposition state. We show that the simple use of a deformable mirror
in a closed-loop configuration allowed to reduce the state noise deriving from phase instabilities.

\begin{figure*}[t]
\centering \includegraphics[width=12cm]{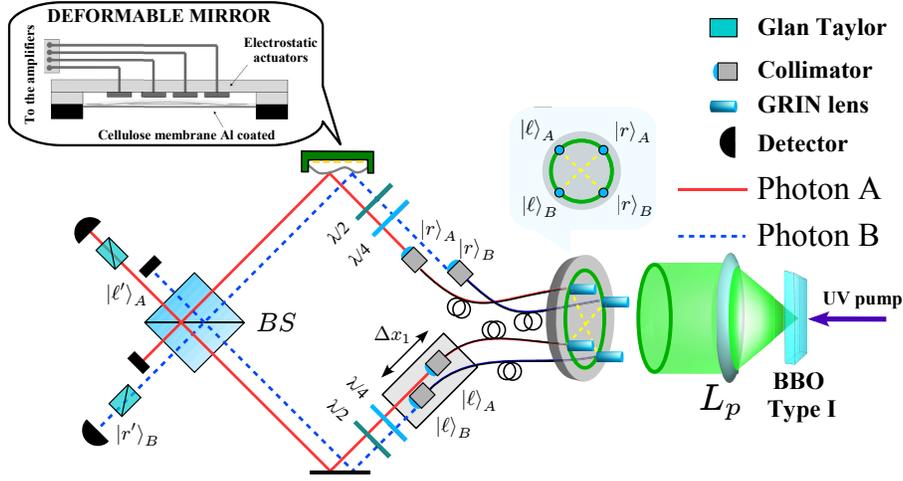}
\caption{Experimental setup. The SPDC source consists of a BBO Type I crystal pumped by a UV laser beam. The parametric radiation, given by four {\bf k} optical modes, is collected by a corresponding number of integrated systems of GRIN lenses and single mode fibers and injected into a two-arm interferometer. Polarization restoration of the photons is performed by proper $\frac{\lambda}{4}$ and $\frac{\lambda}{2}$ wave plate sets after fiber transmission. For each photon, the right ($\ket r$) mode is spatially matched on the BS with the left ($\ket\ell$) mode. A translational stage allows fine adjustment of the left optical paths $\Delta x_1$ to obtain temporal indistinguishability (and thus interference) between the modes. The deformable mirror is placed on the right mode side and allows to change the state phase. Two single photon detectors are placed
after two horizontal (Glan Taylor) polarizers at the output ports of the BS, one on the Alice and the other on the Bob side.}
\label{fig:setup}
\end{figure*}

\begin{figure}[t]
\centering \includegraphics[width=9cm]{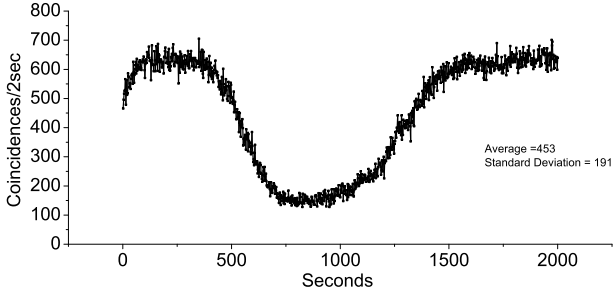}
\caption{Coincidence counts detected by the interferometric setup closed into a thermally isolated box, when no phase control
is activated. Each experimental point represents the number of detected coincidences in 2 seconds. The coincidence rate shows temperature fluctuations within a timescale of the order of hundreds of seconds.}
\label{fig:mirror_off}
\end{figure}

{ Entangling two photons in the longitudinal momentum degree od freedom (DOF), i.e. in different optical paths, is the most efficient way to create quantum states of light spanning high-dimension Hilber spaces, namely qu-dit ($d>2$) and hyper-entangled states. It was already demonstrated that multidimensional entangled states enable the realization of important quantum information tasks,
such as Bell state analysis \cite{schu06prl, barb07pra, wei07pra}, superdense coding \cite{barr08nap},
secure quantum key distribution \cite{bech00prl, thew04qic},
and high fidelity one-way quantum computation \cite{brie01prl, vall07prl, chen07prl, vall08prl}.

Besides other techniques adopted to generate path entanglement, the optical setup used in the present experiment uses photon pairs emitted over the light cone of a nonlinear parametric crystal shined by a laser light. This technique
 may represent a useful resource for an efficient generation and distribution of entangled photon states since it allows one to maximize the emission of photon pairs for a given value of the pump power.

In the experiment, entangled photons are generated with horizontal polarization by Spontaneous Parametric Down Conversion (SPDC) by a $\beta$-borate (BBO) Type-I nonlinear crystal pumped by a continuous wave (cw) ultraviolet (UV) laser beam (wavelength $\lambda_p=266nm$). Couple of photons are emitted time by time at degenerate wavelength $\lambda=2\lambda_p=532nm$ and selected by two interference filters with bandwidth $\Delta\lambda_p=5nm$. Due to momentum conservation, the two photons are emitted with uniform probability distribution along the external surface of a cone, as said, with photon $A$ ($B$) emitted along the up (down) side. A positive lens $L_P$ is then used to transform the conical emission into a cylindrical one. In the present experiment (cf. Fig. ~\ref{fig:setup})) two pairs of opposite correlated directions are collected by four integrated systems, given each by a GRaded INdex (GRIN) lens glued to a single mode fiber \cite{ross09prl, vall09apl}. Then the four integrated systems, pre-aligned to maximize photon coincidences, are glued to a four-hole screen, building in this way a single compact device which 
can be used to study the effects of longitudinal momentum photon entanglement.

The entangled state deriving from the selection of two pairs of SPDC modes is expressed as:
\begin{equation}
\ket{\psi(\phi)}=\frac1{\sqrt2}(\la\rb-e^{i\phi}\ra\lb)
\end{equation}
where $\ket{\ell}$ ($\ket{r}$) refers to the left (right) mode of the corresponding photon \cite{ross09prl}.
Since the two events ${\ket\ell}_A{\ket r}_B$ and ${\ket r}_A{\ket\ell}_B$ assumes the same phase of the laser beam through the SPDC process, their relative phase is due to the differences in fiber and bulk optical paths. Precisely
$\phi=\pi+\frac{2\pi}{\lambda}(r_A+\ell_B-\ell_A-r_B)$, where $\ell_A$ ($\ell_B$) and $r_A$ ($r_B$) are respectively
the left and right path of the photon $A$ ($B$).

The entangled state is injected into the interferometric apparatus shown in Fig.~\ref{fig:setup} where the left and right modes belonging to the $A$ and $B$ modes are mixed on a common beam splitter (BS).
This configuration allows to overcome the mechanical instabilities of the apparatus since any mirror or BS fluctuation affects in the same way both photons and doesn't influence the relative phase of the quantum state. On the other hand the phase $\phi$ is strongly affected by the intrinsic thermal instabilities of the optical fibers.

The BS action on the input modes $\ket{\ell}$ and $\ket{r}$ for both photons can be written as
\begin{equation}
\left\{\begin{aligned}
{\ket{\ell}}_j\rightarrow\frac{1}{\sqrt2}({\ket{\ell'}}_j+i{\ket {r'}}_j)\\
{\ket{r}}_j\rightarrow\frac{1}{\sqrt2}({\ket{r'}}_j+i\ket{\ell'}_j)
\end{aligned}\right.\qquad j=A,B\,,
\end{equation}
where $\ket{\ell'}$ and $\ket{r'}$  are the output modes.

The state corresponding to the BS output is:
\begin{equation}
\begin{split}
\ket{\psi'(\phi)}=\frac1{\sqrt{2}}\left[\frac{1+e^{i\phi}}{2}({\ket{\ell'}}_A{\ket {r'}}_B-{\ket {r'}}_A{\ket{ \ell'}}_B)+\right.\\
\left.i\frac{1-e^{i\phi}}{2}({\ket{\ell'}}_A{\ket {\ell'}}_B+{\ket {r'}}_A{\ket {r'}}_B)\right]
\end{split}
\end{equation}
where $\phi$ depends of the path length difference between $\ra$ and $\rb$. Photon coincidences of the output states are measured by two single photon detectors located each after a Glan Taylor polarizer selecting the horizontal polarization. The rate of a coincidence events at ports ${\ket{\ell'}}_A$ and ${\ket{r'}}_B$ is:
\begin{equation}
\mathcal C(\phi)=
\frac{\mathcal N_0}{4}(1+\cos\phi)
=\frac{\mathcal N_0}{2}\cos^2\frac\phi2
\end{equation}
where $\mathcal N_0$ represents the rate of generated pairs.

In the measurement setup shown in Fig. 1 any temperature variation modifies the optical length of the fibers and hence affctes the phase $\phi$ between the two events $\la\rb$ and $\ra\lb$. In order to minimize the effects of temperature instability the interferometer was placed within a thermally isolated polystirene box. Fig.~\ref{fig:mirror_off} shows the evolution of the coincidence count rate ($2$ $sec$ integration time) over time, with no phase modulation externally applied on the system. The coincidence count rate shows fluctuations over time, with a characteristic time-scale of some tens of seconds. Therefore, the quantum state generated by our system is not stable over time. 

Let's suppose we want to apply a phase shift $\Delta \phi$ to the interferometer in order to create a particular quantum state. The state phase will be expressed as $\phi_0 (t) + \Delta \phi$, where $\phi_0 (t)$ is a stochastic function describing the intrinsic fluctuations of the interferometer. In order to perform a measurement, such $\phi_0 (t)$ should be compensated to zero. In other words, first we need to take the system to a coincidence maximum ($\phi_0 = 0$) and stabilize it, then we can apply the required phase-shift $\Delta \phi$. To take the coincidence rate to a maximum and keep the quantum state stable over time, we compensated phase fluctuations with a deformable mirror placed in one arm of the interferometer and controlling the length difference $\delta$ between the $\ra$ and $\rb$ optical paths. An optical path-length difference $\delta$ corresponds to a phase shift $\phi\rightarrow\phi+{2\pi}\frac{\delta}{\lambda}$, where $\lambda$ is the wavelength of the two photons.}

\begin{figure}[t]
\centering
\includegraphics[width=8cm]{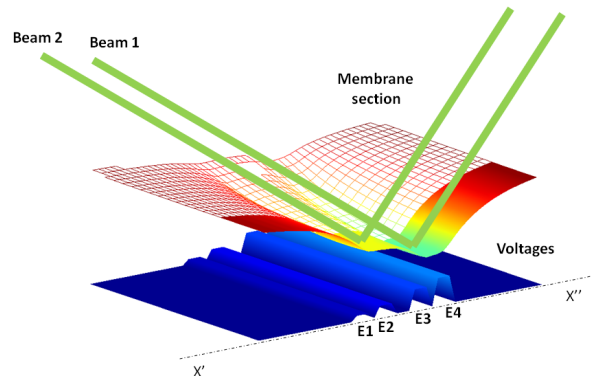}
\caption{Scheme of the deformable mirror}
\label{fig:mirror_scheme}
\end{figure}

\section{Deformable mirror}
A custom deformable mirror was used for phase compensation (see
Fig.~\ref{fig:mirror_scheme}). It basically consists  of an
aluminized nitrocellulose membrane which is deflected by the
electrostatic pressure applied trough a series of pads placed
hundred microns below the membrane \cite{bono06rsi}. The electrodes
were controlled by a high voltage (0-265V) driver which can
independently address the actuators. Such deformable mirrors are
usually used for aberration compensation \cite{fern03ope} or, in
some cases, for the compression of ultrafast pulses\cite{brid08opl}.
The mirror design was such that two square areas (size 1.4mm by
1.4mm) of the membrane behave like flat, parallel, mirrors with
controllable relative displacement \emph{d} (see Fig.~\ref{fig:mirror_scheme}). This
allowed us to control the relative phase shift of the two photon
beams, which were spatially few millimeters close (beam diameter
around $1$ mm).

The best membrane shape for carrying on this task is rectangular
because the membrane boundaries are parallel to the planes. Because
of the strong crosstalk between the deformation caused by the single
electrodes we had to compute the square areas position which allow a
large enough \emph{d} displacement keeping the flatness and
parallelism of the planes suitable for the experiment. A preliminary
study of the deformation $M(x,y)$ was carried out solving the
Poisson equation for membranes under an applied voltage \cite{claf86jos}:
\begin{equation}
\nabla M(x,y) = - \frac{1}{T}p(x,y),
\end{equation}
where $p$ is the electrostatic pressure,
\begin{equation}
p(x,y) = \frac {\epsilon_{0}}{2}(\frac{V(x,y)}{h})^2
\end{equation}
and T is the mechanical tension of the membrane and $h$ the membrane
to electrodes distance. The simulations were carried out through
Finite Element Method which allowed to design the deformable mirror.
In order to address such a membrane deformation we used two couple
of electrodes with the size of $1.4$ mm by $15$ mm spaced of $1.4$
mm. The electrodes are empty in the central part in order to
completely remove the transverse radius of curvature
\cite{bono08spie}. In order to compute the voltages necessary for
controlling the membrane deformation we measured the shape obtained
applying the maximum voltage to each actuator (influence function)
using an interferometric technique. Under the hypothesis of
linearity, valid for small membrane deformations, we combined the
influence functions to determine the voltages that create the two
parallel planes with a controllable displacement \emph{d} minimizing
the root-mean-square (\emph{rms}) flatness error. To compute the
position of the planes we used the following procedure: the position
of plane 1 was kept fixed and the position of plane 2 was increased
until the rms deviation from a plane parallel to a reference was
smaller than a threshold value of 30nm rms. Then we repeated this
algorithm increasing the position of plane 1. Following this
strategy in first instance we determined the optimal plane distance
which was 11.2mm. Then we characterized the performances of the
deformable mirror as phase shifter. The maximum displacement
achievable was about 600nm with 8bit control resolution. We measured
over the whole displacement range the average rms deviation from the
reference plane which was of $25.98$nm for the fist plane and
$23.77$nm for the second. Moreover we measured the parallelism of
the two planes which was within $17.66 \pm 4.3\mu$rad.

\begin{figure}[t]
\centering \includegraphics[width=7cm]{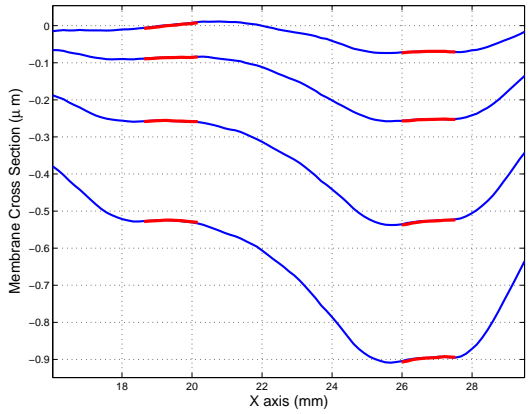}
\caption{Measurements of the plot of the cross section X'-X'' of the membrane
for four different values of relative displacement.
The portions of the membrane used for phase shifting of the two
beams is in red color. The measurement was carried out with
interferometric technique.}
\label{fig:mirr_deform}
\end{figure}

The deformation of the membrane for different applied voltages is illustrated in Fig~\ref{fig:mirr_deform}. The red curve shows the shape of the membrane, where two flat portions in blue are used for the relative shift of the two beams.

\begin{figure}[t]
\centering \includegraphics[width=9cm]{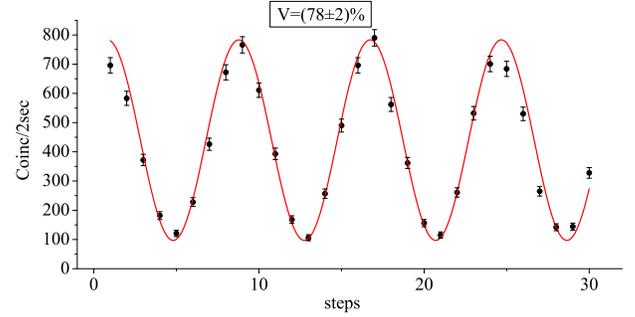}
\caption{Mirror calibration. Measurement of coincidence counts as a function of mirror deformation.
This allow to calibrate the deformation in terms of the state phase. In the graph each step corresponds to a $\pi/4$ phase shift.
}
\label{fig:oscillation}
\end{figure}

\begin{figure}[t]
\centering
\includegraphics[width=9cm]{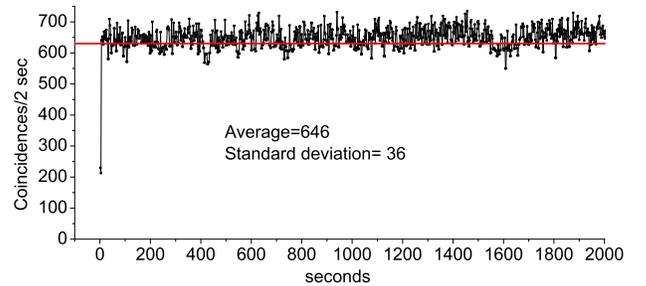}
\caption{Coincidence counts with activated deformable mirror.
The first point represents the
initial random phase. The optimization algorithm rapidly controls the phase state to maximize the coincidences and to keep
the phase constant to $\phi=0$.}
\label{fig:mirror_on}\end{figure}

{\section{Measurements}}
Before applying a control algorithm, the system was characterized scanning $\delta$ with the deformable mirror, over a time scale much shorter than the instability time. Starting from a coincidence maximum, $\delta$ was increased by steps of $\pi/4$ ($30$ steps in total, over a $60$ s time). The coincidence count rates exhibits a sinusoidal behavior as a function of the externally applied phase (see Fig.~\ref{fig:oscillation}), with a maximum of $800$ and a minimum of $100$ coincidences in $2$ seconds (visibility $78 \pm 2 \%$).

In order to select a proper phase $\phi$ for the quantum state we assume to take the coincidence rate to the maximum, which we set as $\phi_0 = 0$, an then apply the needed phase-shift $\phi$. Therefore, our problem can be reduced to the one of taking the interferometer into the $\phi_0 = 0$ state, maximizing the number of coincidences. From the preliminary characterization  of the system parameters, we assume to know the expected value of coincidences per second of the maximum $\mathcal{C}_{max}$.

\begin{figure}[t]
\centering
\includegraphics[width=8.7cm]{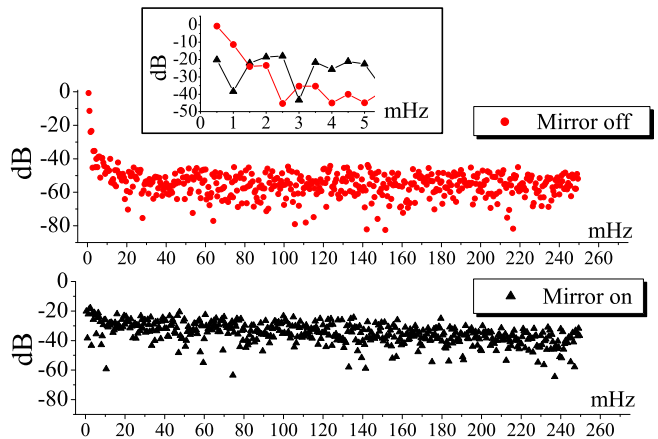}
\caption{
Discrete Fourier transform of the data shown in Fig. \ref{fig:mirror_off} and \ref{fig:mirror_on}.
The Fourier components are normalized such that the sum of their squares is equal to 1.
The frequency components at $0.5mHz$ and $1mHz$ (see inset) are filtered out by the action of the deformable mirror.}
\label{fig:fourier}\end{figure}

In a typical closed-loop experiment, a measurement is performed on the system at each step and the measurement result is used as a parameter to drive the controller. In our case, we estimate the phase of the quantum state measuring the coincidence rate and comparing it to its maximum value. To remove the phase ambiguity (different phase values giving the same coincidence rate) we compare this value with the coincidence rate obtained increasing the phase by a small amount. In other words, we estimate the phase $\phi$ of the quantum state by measuring the coincidence rate $\mathcal{C} (\phi) $ and its derivative with respect to phase $d\mathcal{C}(\phi)/d\phi$. Then we use this estimate of $\phi$ to guess what's the phase we need to apply in order to take the system to the maximum. We take a threshold value $T$, as the value above which the coincidence rate can be assumed to be at the maximum. For example, we can take the threshold to be one standard deviation below the expect maximum value: $T = \mathcal{C}_{max} -\sqrt{\mathcal{C}_{max}}$. In details, the maximization algorithm we propose works as follow:
\begin{enumerate}
\item if the coincidence rate $\mathcal C$ is above $T/2$ and below $T$, then we can apply exactly the phase shift we need to get to the closest maximum: $\Delta \varphi = 2\arccos\sqrt{\frac{\mathcal C}{T}}$. Knowing only $\mathcal C$ we cannot determine the sign of $\Delta \varphi$, since we could be either in the ascending or descending side of the maximum. Therefore we use a double-step procedure: to determine in which side of the maximum we are, we apply a small phase shift. If the number of coincidences increases, we are on the ascending side and we apply $+\Delta \varphi$. If the number of coincidences decreases, then we apply $-\Delta \varphi$.
\item if the coincidence rate is below $T/2$, then we take the system into case (1), shifting the phase by $\pi$.
\item Above the threshold $T/2$ the maximization procedure is successful and the mirror can be switched off.
\end{enumerate}
The results shown in Fig.~\ref{fig:mirror_on} demonstrate that the deformable mirror can compensate the slow temperature fluctuations causing the coincidence variation given in Fig. \ref{fig:mirror_off}. The maximization procedure rapidly converges to a coincidence value above the threshold.
The standard deviation of the stabilized data ($\sigma=36$) is comparable with the Poissonian error of the average coincidence, namely $\sigma_P=\sqrt{646}\simeq25$ ($\sigma = 1.44 \sigma_P$).
In Fig.~\ref{fig:fourier} the discrete Fourier transform of the data with and without adaptive compensation is plotted. The intensity of the frequency components below $1mHz$, due to the slow phase fluctuations shown in Fig.~\ref{fig:mirror_off}, is clearly reduced.

The mirror can now be used to fix an arbitrary phase state. By accurate calibration we can obtain the precise phase state variation in terms of the mirror
deformation. Thus, after a preliminary stabilization with $\phi=0$ the mirror can be used to generate a state with arbitrary phase $\phi$.

It is worth to note that the control scheme we proposed is different from quantum feedback control. In feedback quantum control, a single quantum system is subject to control to force its dynamics according to some requirements. Measurements, used to provide input information for the control, are quantum: the result is probabilistic and the measurement process itself has a back-action on the state, in the sense that it projects the  state to one of the eigenvectors of the measurement operator.  On the other hand, in our case the measurements we are performing on the system are in a sense 'classical', since we repeat the experiment on different copies of the same input quantum state traveling through the optical interferometric system, having access to the mean value of the result. We do not have a single quantum state that evolves under the influence of the external environment. We have a source, which emits a stream of quantum states which evolves in an unpredictable way over time, and we want to keep it stable. Moreover, in photon-counting experiments, the currently available photo-detectors absorb photons, so that each single quantum state produced is destroyed in the measurement process and cannot be used for further operations. 

Another important aspect of our experiment regards the involved time-scales. In photon counting experiments the number of counts is a Poissonian process and a sufficient amount of counts $N$ needs to be collected in order to have a good signal-to-noise ratio (which scales like $1/\sqrt{N}$). Entangled photon sources based on SPDC typically provide a few thousand pairs per second, which means that to reach a signal-to-noise ratio around $1\%$ one needs to collect counts for at least one second. Since the measurement time must be much faster than the fluctuations, this sets a bound on the time-scale of the fluctuations that can be compensated.

In conclusion, we experimentally demonstrated a feedback control of the states generated by a source of entangled photon pairs by means of a custom-design deformable mirror. We believe this technique can be extremely beneficial to quantum interference experiments, since it decouples the quantum state produced by the source from the random phase fluctuation induced by the environment.

\end{document}